\documentstyle[prb,aps,psfig]{revtex}

\begin{document}
\draft
\twocolumn[\hsize\textwidth\columnwidth\hsize
\csname@twocolumnfalse\endcsname

\title{Coexistence of double alternating antiferromagnetic chains\\ in 
(VO)$_2$P$_2$O$_7$ : NMR study}

\author{Jun Kikuchi,$^{*}$ Kiyoichiro Motoya}
\address{Department of Physics, Faculty of Science and Technology,
Science University of Tokyo,\\ 2641 Yamazaki, Noda, Chiba 278-8510, Japan} 
\author{Touru Yamauchi and Yutaka Ueda} 
\address{Institute for Solid State Physics, University of Tokyo,\\
7-22-1 Roppongi, Minato-ku, Tokyo 106-8666, Japan} 

\date{\today} 
\maketitle

\begin{abstract}
   Nuclear magnetic resonance (NMR) of $^{31}$P and $^{51}$V nuclei has 
been measured in a spin-1/2 alternating-chain compound (VO)$_2$P$_2$O$_7$. 
By analyzing the temperature variation of the 
$^{31}$P NMR spectra, we have found that (VO)$_2$P$_2$O$_7$ has two 
independent spin components with different spin-gap energies.  The 
spin gaps are determined from the temperature dependence of the shifts 
at $^{31}$P and $^{51}$V sites to be 35 K and 68 K, which are in 
excellent agreement with those observed in the recent inelastic neutron 
scattering experiments {[}A.W.  Garrett {\it et al}., Phys.  Rev.  
Lett.  {\bf 79}, 745 (1997){]}.  This suggests that (VO)$_2$P$_2$O$_7$ 
is composed of two magnetic subsystems showing distinct magnetic 
excitations, which are associated with the two 
crystallographically-inequivalent V chains running along the $b$ 
axis.  The difference of the spin-gap energies between the chains is 
attributed to the small differences in the V-V distances, which may 
result in the different exchange alternation in each magnetic chain.  
The exchange interactions in each alternating chain are estimated and 
are discussed based on the empirical relation between the exchange 
interaction and the interatomic distance.

\end{abstract}


]

\narrowtext

\section{Introduction}
\label{sec:intro}

   Low-dimensional quantum-spin systems with antiferromagnetic 
interactions exhibit various interesting phenomena of collective nature.  
Vanadyl pyrophosphate (VO)$_2$P$_2$O$_7$ is a gapped antiferromagnet with 
spin 1/2 and had long been known as a good model compound of a two-leg 
antiferromagnetic Heisenberg spin 
ladder.\cite{johnston87,barnes94,eccleston94} Recently, dispersion of 
the magnetic excitations was measured using the inelastic 
neutron-scattering technique and is found to be inconsistent with a 
simple ladder model.\cite{garrett97} Although an alternating 
Heisenberg antiferromagnetic chain is proposed as an alternative 
magnetic model of (VO)$_2$P$_2$O$_7$, the suggested direction of the 
chain is not readily known from the crystal structure and the 
excitation spectrum is not fully compatible with a simple alternating 
spin chain.  It is still a controversy what magnetic model describes 
best the magnetic behavior of (VO)$_2$P$_2$O$_7$.

\begin{figure}
\centerline{\psfig{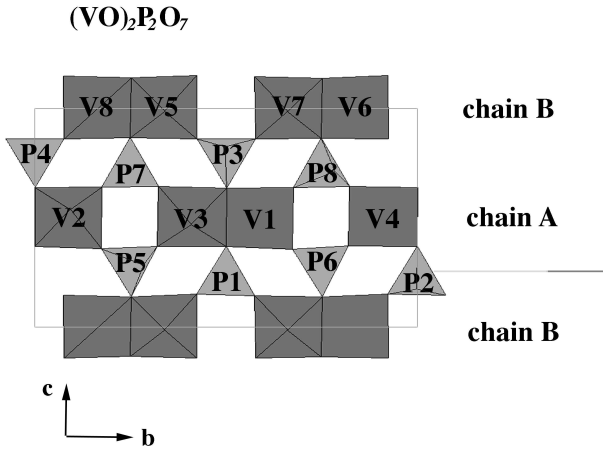}}
\medskip 
\caption{
Polyhedral view of the crystal structure of (VO)$_2$P$_2$O$_7$.  
Also indicated are site indices of the inequivalent V and P atoms included 
in each polyhedron. The structural units indicated are stacked in the $a$ 
direction. See text for the labeling of the chains 
directed along the $b$ axis.}
\label{structure}
\end{figure}

   The ladder interpretation of (VO)$_2$P$_2$O$_7$ is based on the intuitive 
consideration of the crystal structure which is shown schematically in Fig.\ 
\ref{structure}.  (VO)$_2$P$_2$O$_7$ belongs to the monoclinic space 
group $P$2$_1$ and has unit-cell dimensions of $a$ = 7.28 {\AA}, $b$ = 
16.59 {\AA}, $c$ = 9.58 {\AA} and $\beta $ = 89.98$^\circ$ at room 
temperature.\cite{nguyen95} The vanadium atoms are tetravalent to have 
spin-1/2 degrees of freedom, each of which is coordinated pyramidally 
by oxygen atoms.  The VO$_5$ pyramids form pairs in opposite phases in 
the $b$ direction by sharing their edges, and are stacked in pairs 
along the $a$ axis to make up a two-leg ladder.  The ladders are 
linked by PO$_4$ tetrahedra and form a three-dimensional lattice 
structure.  The proposed direction of the alternating chain is along 
the $b$ axis which is perpendicular to the ladder axis and the chain 
includes PO$_4$ tetrahedra as one of the important exchange 
pathways.\cite{garrett97} There are eight inequivalent V and P sites 
in the unit cell, so that the ladders and chains are not all 
crystallographically equivalent.

   The magnetic susceptibility of (VO)$_2$P$_2$O$_7$ has a broad maximum 
around 80 K and decreases rapidly at low temperatures.  Johnston {\it et al}.  
have pointed out that (VO)$_2$P$_2$O$_7$ has a spin-singlet ground state 
owing to the ladder configuration of magnetic atoms, although the spin-1/2 
alternating Heisenberg antiferromagnetic chains described by the 
Hamiltonian
\begin{equation}
\label{alternate}
	{\cal H} = \sum_{i} (J_{1} {\bf S}_{2i} \cdot {\bf S}_{2i+1} +
J_{2} {\bf S}_{2i+1} \cdot {\bf S}_{2i+2}).
\end{equation}
accurately reproduces the susceptibility of (VO)$_2$P$_2$O$_7$.
\cite{johnston87} Numerical calculations of the susceptibility of both the 
two-leg ladder and the alternating chain showed that both models fit the 
susceptibility of (VO)$_2$P$_2$O$_7$ quite well, so that the susceptibility 
measurement alone cannot distinguish which model is appropriate.\cite{barnes94} 
The calculations also showed that the magnitude of the spin gap is different 
depending on the models, which can be tested by inelastic neutron scattering 
(INS) experiments.  The subsequent INS measurement on powder sample of 
(VO)$_2$P$_2$O$_7$ has detected a spin gap of 3.7 meV, which favors 
the ladder model.\cite{eccleston94}

   More recently, however, Garrett {\it et al}. have presented a  
surprising INS result on the magnetic excitations from an array of a lot of 
small single crystals.\cite{garrett97} They found that the spin excitation is 
predominantly one dimensional but is directed along the crystalline 
$b$ axis rather than the ladder (the crystalline $a$) axis.  The 
exchange interaction along the $a$ axis is ferromagnetic and is order 
of magnitude smaller than that along the $b$ axis, and there is 
essentially no dispersion along the $c$ axis.  As the most likely 
model to account for the INS result, they proposed that (VO)$_2$P$_2$O$_7$ has 
alternating antiferromagnetic chains running in the $b$ direction with 
unexpectedly strong V-V exchange via the PO$_4$ complexes.  The exchange 
interaction $J_1$ via the PO$_4$ complexes is as large as 10$\sim$12 
meV and the alternation ratio $J_2/J_1$ is estimated to be $\sim$0.8.  
A possible importance of the superexchange pathways through the PO$_4$ 
complexes was pointed out nearly a decade ago by Beltr\'{a}n-Porter 
{\it et al}.  to account for the magnetic properties of various 
vanadyl phosphates,\cite{beltran89} and has recently been observed by 
INS in the dimer counterpart VODPO$_4$$\cdot$${1\over 
2}$D$_2$O.\cite{tennant97}

   Another important but unresolved finding of Garrett {\it et al}. is the 
existence of a second excitation mode whose zone-center gap is 5.7 meV. 
They associated it with the two-magnon bound state as it lies just below 
2$\Delta$ ($\Delta$ = 3.1 meV being the gap for the one-magnon excitation 
determined by Garrett {\it et al}.).  However, as they have pointed out, the 
substantial intensity of the second mode is difficult to interpret for the 
two-magnon bound state. Frustrated interchain coupling has recently been 
suggested to give rise to a triplet two-magnon bound state just below 
2$\Delta$ which is observable over a large part of the Brillouin 
zone.\cite{uhrig98} However, the model is rather complicated and 
cannot reproduce correctly the intensity of the second mode especially 
near the zone center, where the distinct peak of the second mode is 
observed opposed to the theoretical prediction.

   In this paper, we report on the results of $^{31}$P and $^{51}$V NMR on 
high-quality (VO)$_2$P$_2$O$_7$ powder.  We have found two independent 
magnetic subsystems existing in (VO)$_2$P$_2$O$_7$ whose spin gaps are 
exactly the same as those of two distinct modes in the spin-excitation 
spectrum.  The existence of two magnetic subsystems with different 
spin gaps is supported by the recent high-field magnetization 
measurement.\cite{yamauchi99} From crystallographic considerations and 
a possible correlation between the exchange interaction and the 
interatomic distance, we suggest that (VO)$_2$P$_2$O$_7$ has two 
inequivalent alternating antiferromagnetic chains directed along the 
$b$ axis.  The role of the PO$_4$ complexes as the exchange media and 
in determining the microscopic magnetic properties of 
(VO)$_2$P$_2$O$_7$ will also be discussed in detail.

   This paper is organized as follows. We start with describing experimental 
details in Sec.\ \ref{sec:expt}. The results of the NMR measurements will be 
presented in Sec.\ \ref{sec:results}. The analysis of the line shape, the 
shifts and the nuclear spin-lattice relaxation rates will also be given in 
Sec.\ \ref{sec:results}.  Section\ \ref{sec:discussion} is devoted to the 
discussion of the experimental results and their interpretations, followed by 
a conclusion drawn from the present experimental 
findings in Sec.\ \ref{sec:conclusion}.

\section{Experimental}
\label{sec:expt}

   Polycrystalline sample of (VO)$_2$P$_2$O$_7$ is prepared by following 
several steps of reaction.  The first step is to synthesize 
VOPO$_4$$\cdot$2H$_2$O as a precursor by intercalating V$_2$O$_5$ with 
H$_3$PO$_4$ in ethanol at room temperature for about a month.  The 
precursor is then heated in air at 650 $^\circ$C for 2 days to remove 
water completely.  The final product (VO)$_2$P$_2$O$_7$ is 
obtained by reducing oxygen from VOPO$_4$ in Ar atmosphere at 700 
$^\circ$C for 2 days.  The X-ray diffraction pattern of the obtained 
sample indicates that there is no trace of impurity phases. NMR 
measurements were performed with a standard phase coherent-type pulsed 
spectrometer.  The $^{31}$P NMR spectra were taken by integrating the 
spin-echo signal with a box-car averager while sweeping the external magnetic 
field at a fixed frequency.  The $^{51}$V NMR spectra were obtained by 
Fourier transforming the free-induction-decay signal after a $\pi/2$ 
pulse at a fixed frequency and a magnetic field.  Nuclear spin-lattice 
relaxation rates were measured by the saturation recovery method with 
a single saturation rf pulse for both the $^{31}$P and $^{51}$V sites.

\section{Results and analysis}
\label{sec:results}
\subsection{$^{31}$P NMR}
\subsubsection{Line shape and the shifts}

\begin{figure}
\centerline{\psfig{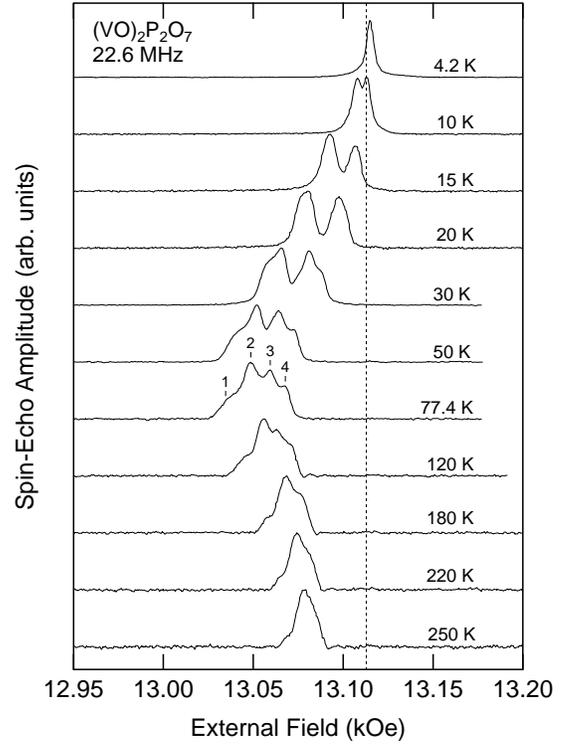}}
\medskip 
\caption{
Temperature variation of the $^{31}$P NMR spectrum taken 
at 22.6 MHz.  Dashed line indicates the position of the zero shift 
($K$ = 0) for $^{31}$P nuclei calculated using the nuclear gyromagnetic 
ratio $^{31}$$\gamma$$_n$ = 17.235 MHz/T.  Numbering of the peaks (see 
text) is shown for the spectrum at 77.4 K.}
\label{spectra}
\end{figure}

   The field-swept $^{31}$P NMR spectra taken at various temperatures are 
shown in Fig.\ \ref{spectra}.  Several peaks are found in the spectrum as 
expected from the crystal structure, and the line shape is strongly 
temperature dependent.  At 4.2 K, a sharp single peak was observed near the 
zero shift $K$ = 0, indicating the nonmagnetic ground state of 
(VO)$_2$P$_2$O$_7$.  The width of the spectrum is very narrow ($\sim$ 4 Oe) and 
is attributed to the broadening due to P nuclear dipoles.  The absence of 
inhomogeneous broadening of the spectrum assures the homogeneity of the sample 
in a microscopic scale and the extremely low concentration of paramagnetic 
impurities.  As the temperature increases, the spectrum splits into two peaks 
at first, and each peak further splits to show fine structure. Such a fine 
structure was not observed in the previous reports probably because of 
inhomogeneous broadening.\cite{furukawa96,kikuchi97} The multi-peak 
structure can be seen most clearly around 80 K where the resonance 
lines are most distant from the $K$ = 0 position.  There are four 
peaks visible in this temperature region, indicating that there exists 
at least four distinct P sites which feel different local magnetic fields from 
surrounding V atoms.  Note that the number of the observed peaks is 
less than the number of the crystallographically-inequivalent P sites.  
This suggests that the site distinction among some P sites is not 
clear and is indistinguishable in the present experiment.  In the 
following, the observed peaks are referred to as the peaks 1 to 4 from 
low-field side to the higher.  As the temperature is increased 
further, the fine structure is obscured and the spectrum merges into a 
single peak again (although the structure is visible at 250 K).

\begin{figure}
\centerline{\psfig{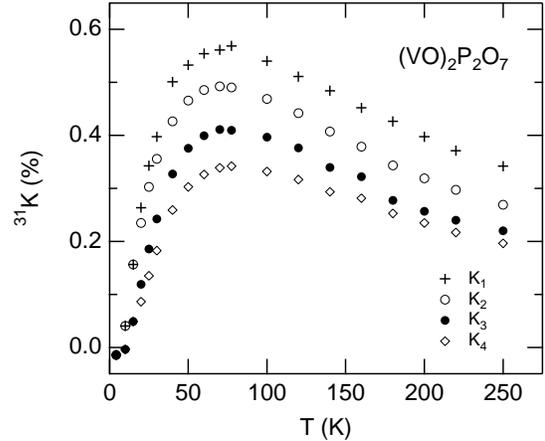}}
\medskip
\caption{Temperature dependence of the shifts at $^{31}$P sites.}
\label{K31vsT}
\end{figure}

   The nonmonotonous variation of the line shape with temperature suggests 
the existence of several spin components having different spin susceptibility 
($\chi_{\rm spin}$).  To trace the temperature ($T$) dependence of each spin 
component, we deconvoluted the spectrum into four gaussians and 
determined the shift of each peak $K_i$ ($i$ = 1, 2, 3, 4).  The 
results are shown in Fig.\ \ref{K31vsT} where $K_i$'s are plotted as a 
function of temperature.  The $T$ dependence of the $K_i$'s is 
distinct from each other in magnitude but is similar in that they all 
exhibit a broad maximum around 80 K corresponding to the bulk magnetic 
susceptibility.\cite{similarJ} At low temperatures, all the $K_i$'s 
decrease rapidly to almost zero owing to the gap in the spin-excitation 
spectrum.

\begin{figure}
\centerline{\psfig{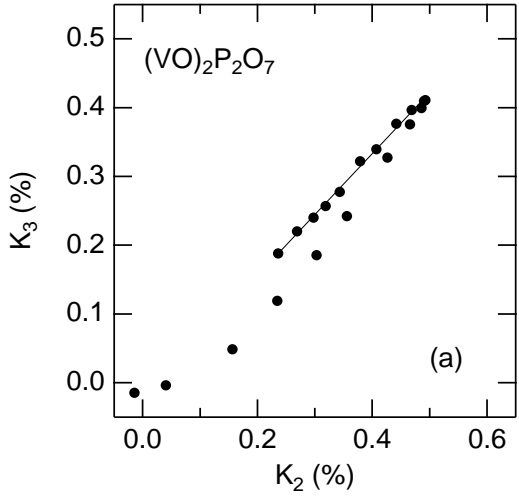}}
\medskip
\centerline{\psfig{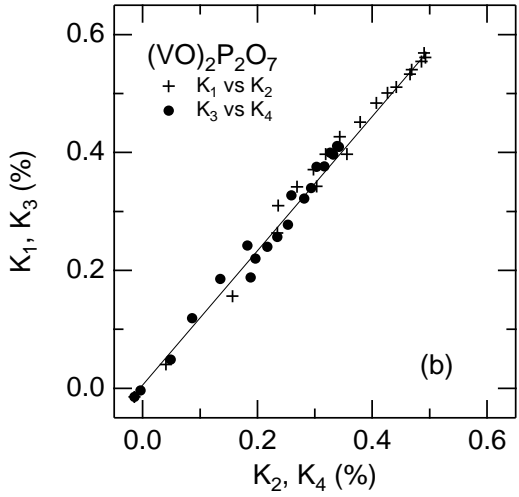}}
\medskip
\caption{
(a) The shift of the peak 3 ($K_3$) plotted against the shift of 
the peak 2 ($K_2$) with temperature the implicit parameter.  The straight line 
is a fit of the data from 100 to 250 K.  (b) Plots of $K_1$ versus $K_2$ and 
$K_3$ versus $K_4$.  The line is a guide to the eyes of which slope 
corresponding to the average of $A_1/A_2$ and $A_3/A_4$.}
\label{KvsK}
\end{figure}

   The difference of the shifts at the crystallographically-inequivalent 
nuclear sites is usually attributed to the difference of the hyperfine 
coupling constants if there is only one kind of spin component.  
However, this is not the case in (VO)$_2$P$_2$O$_7$.  Figure\ 
\ref{KvsK}(a) shows the shift of the peak 3 ($K_3$) plotted against 
the shift of the peak 2 ($K_2$) with temperature the implicit 
parameter.  If the $T$ dependence of $K_2$ and $K_3$ is the same, the 
$K_3$ versus $K_2$ plot should lie on a straight line and its slope 
gives the ratio of the coupling constants $A_3/A_2$ ($A_i$ being the 
hyperfine coupling constant at P sites associated with the peak $i$).  
It is obvious from Fig.\ \ref{KvsK}(a) that $K_3$ does not scale to 
$K_2$ at low temperatures (near the origin of the $K_3$-$K_2$ plot).  
There the $K_3$ versus $K_2$ plot concaves downward, which indicates 
that $K_2$ increases more rapidly with increasing temperature than 
$K_3$.  This suggests that there are at least two independent spin 
components in (VO)$_2$P$_2$O$_7$ whose $\chi_{\rm spin}$'s are 
mutually different.

   It is interesting to note that at relatively high temperatures, a nice 
linear relation is found between $K_3$ and $K_2$.  This indicates that the 
$T$ dependence of $\chi_{\rm spin}$'s responsible for 
$K_3$ and $K_2$ is almost identical in this temperature region.  This is not 
surprising because at temperatures much higher than the exchange 
interactions, spins behave paramagnetically and the asymptotic 
relation $\chi_{\rm spin} \approx C/T$ with $C$ the Curie constant for 
$S$ = 1/2 is expected to hold for all the spin components.  From the 
slope of this linear portion, the ratio of the hyperfine coupling 
constants $A_3/A_2$ is determined to be 0.88 $\pm$ 0.02.

   We have made a similar analysis for all possible pairs of $K_i$'s and have 
found that the observed four peaks can be divided into two groups.  These are 
the peaks 1 and 2 (group I) and the peaks 3 and 4 (group II) each of which 
shows distinct $T$ dependence of the shifts.  This indicates that there 
exist two independent spin components in (VO)$_2$P$_2$O$_7$. 
Figure\ \ref{KvsK}(b) shows the plots of $K_1$ versus $K_2$ and $K_3$ versus 
$K_4$ similar to Fig.\ \ref{KvsK}(a). A linear relation is found between 
$K_1$  ($K_3$) and $K_2$ ($K_4$), indicating that their $T$ dependences are 
identical aside from the $T$-independent prefactors (or $A_i$'s).  It is 
surprising that both plots lie on almost the same straight line in spite of 
the difference in the $T$ dependence between the groups.  Indeed, the ratios 
of the coupling constants $A_1/A_2 = 1.15 \pm 0.03$ and $A_3/A_4 = 1.18 
\pm 0.04$ agree within experimental accuracies.  This suggests that both P 
groups have similar local environment with regard to the site 
distinction among P sites within the same group.  It is also noted that the 
intensity ratio of the two different P groups is close to unity ($I_{\rm 
I}/I_{\rm II} = 1.1 \pm 0.1$).  This indicates that the number of 
nuclei contributing to the NMR line of each P group is almost the same.

\begin{figure}
\centerline{\psfig{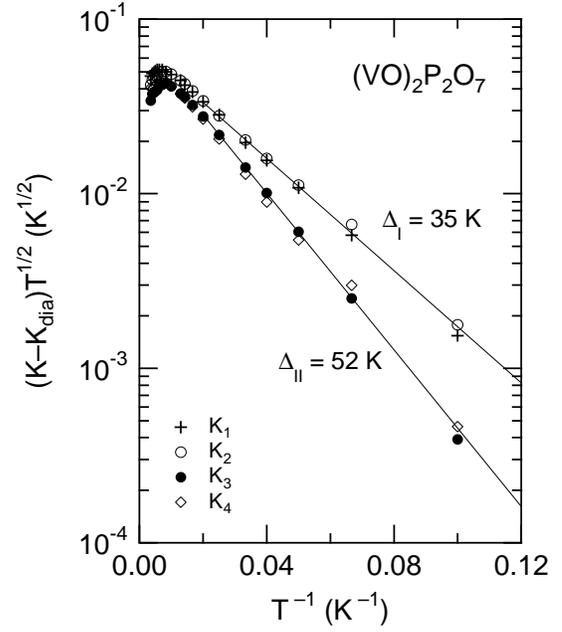}}
\medskip
\caption{
Plots of $(K-K_{\rm dia})T^{1/2}$ against $1/T$.  The data 
for $K_1$ and $K_4$ are plotted by multiplying the scaling factors 
$A_2/A_1$ and $A_4/A_3$ to absorb the difference of the hyperfine 
coupling constants within the groups.  The lines indicate the 
activation law $(K-K_{\rm dia})T^{1/2}\propto \exp (-\Delta /T)$ with 
$\Delta$ of 35 K and 52 K for the groups I and II, respectively.}
\label{K31actv}
\end{figure}

   As demonstrated in Fig.\ \ref{KvsK}(a), the difference in the 
$T$ dependence of the shifts at P sites belonging to the 
different groups is evident at low temperatures. This implies that the 
magnitude of the spin gap is different for the individual spin 
components.  To estimate the gap of each spin component, we analyzed 
the $K_i$'s based on the susceptibility of a gapped one-dimensional 
spin system.\cite{sachdev97,damle98} If the magnon dispersion along the 
chain is approximated by the quadratic form 
$\epsilon(k)\simeq\Delta+c^2k^2/2\Delta$ near the bottom of the 
dispersion ($k=q-\pi \sim 0$), the $T$ dependence of the 
susceptibility $\chi$ in the low-temperature limit $T$ $\ll$ $\Delta$ 
is expressed as
\begin{equation}
\label{Xspin}
	\chi = \sqrt {2\Delta \over \pi c^{2}T} e^{-\Delta /T}
\end{equation}
where $c$ is the spin velocity.  We fitted the $K_2$ 
(group I) and $K_3$ (group II) data below 25 K to the form
$K=K_{\rm dia}+\alpha T^{-1/2}\exp (-\Delta /T)$
where $K_{\rm dia}$ is a $T$-independent diamagnetic 
contribution to the shifts.  The obtained values are $K_{\rm dia} = 
-0.016 \pm 0.005$ \% for both $K_2$ and $K_3$, $\alpha = 0.065 \pm 
0.002$ K$^{1/2}$ and $\Delta = 35 \pm 2$ K for $K_2$, $\alpha = 0.079 
\pm 0.002$ K$^{1/2}$ and $\Delta = 52 \pm 3$ K for $K_3$.  The results 
of the fit are shown in Fig.\ \ref{K31actv} where we plotted $(K-K_{\rm 
dia})T^{1/2}$ against $1/T$.  We also plotted in the same figure 
the corresponding values of $K_1$ and $K_4$ with the appropriate 
scaling factors (i.e.  the ratios $A_1/A_2$ and $A_3/A_4$ 
determined above) and $K_{\rm dia}$.  The correspondence between $K_1$ 
($K_3$) and $K_2$ ($K_4$) is quite well. It should be noted that the estimated 
gap $\Delta_{\rm I}$ = 35 K for the group I agrees within the experimental 
accuracy with the gap of the lower mode (36 K) observed in the INS  
on single crystals, while the gap $\Delta_{\rm II}$ = 52 K for the group II is 
somewhat smaller than the gap for the upper mode (67 K).

   The observed two independent spin components in (VO)$_2$P$_2$O$_7$ may be 
associated with the two crystallographically-inequivalent magnetic chains 
running along the $b$ axis. The two chains have different alternation in the 
V-V distances, so that the gap may be different from each other. This would 
give a natural explanation for the existence of the second mode in the 
excitation spectrum, i.e., the second mode is not the two-magnon bound 
state but the independent magnetic excitation from another alternating 
antiferromagnetic chain with a larger spin gap.  The fact that the NMR 
lines for the groups I and II have almost the same intensity supports 
this interpretation.

   If we assume that the $T$ dependence of $\chi_{\rm spin}$'s of the 
two spin components is identical at high temperatures which is 
justified by the presence of a linear region in the $K_3$ versus $K_2$ 
plot, the absolute values of the $A_i$'s can be estimated from the slopes 
of the usual $K$-$\chi$ plots.  The plots of the $K_i$'s against 
$\chi$ all have linear regions at high temperatures, the slopes of 
which yield $A_1 = 20 \pm 2$ kOe/$\mu_{\rm B}$, $A_2 = 17 \pm 1$ 
kOe/$\mu_{\rm B}$, $A_3 = 15 \pm 1$ kOe/$\mu_{\rm B}$ and $A_4 = 13 
\pm 1$ kOe/$\mu_{\rm B}$.  These values give the ratios $A_3/A_2$, 
$A_1/A_2$ and $A_3/A_4$ consistent with those determined independently 
from the $K$-$K$ scaling shown in Fig.\ \ref{KvsK}.

\subsubsection{Nuclear spin-lattice relaxation rates}

   The existence of two independent spin components with different energy 
gaps in (VO)$_2$P$_2$O$_7$ will be demonstrated more directly if the nuclear 
spin-lattice relaxation rates $1/T_1$ for the different P groups can be 
measured separately.  However, the peaks overlap severely in the spectrum over 
a wide temperature range, which makes independent determination of 
$1/T_1$ rather difficult.  We therefore measured $1/T_1$ in the 
limited temperature range from 12.5 to 30 K where the peaks for the 
different P groups are well articulated, and have successfully 
determined $1/T_1$ for the groups I and II.

\begin{figure}
\centerline{\psfig{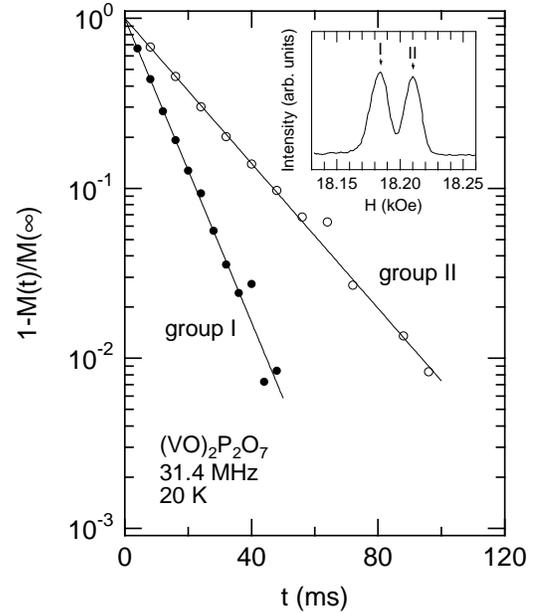}}
\medskip
\caption{
Typical examples of the recovery curve of $^{31}$P nuclear 
magnetization.  The data were taken at 31.4 MHz and 20 K.  Inset shows the 
field-swept $^{31}$P NMR spectrum at the same frequency and 
temperature.  Arrows in the inset indicate the positions of the external 
fields where the recovery curves are measured.}
\label{recovery}
\end{figure}

   The recovery curves were taken at the peak position of each resonance 
line. To avoid excitations of nuclear spins belonging to the different group, 
we used a relatively small rf exciting field $H_1$.  A typical strength 
of $H_1$ was 14 Oe which exceeds the width of each peak and is still 
smaller than the line splitting (25-30 Oe at 31.4 MHz where the 
measurements were performed).  The recovery curves can be fitted to an 
exponential function as illustrated in Fig.\ \ref{recovery}, 
indicating that we reliably separate the signal from the different P 
group.

\begin{figure}
\centerline{\psfig{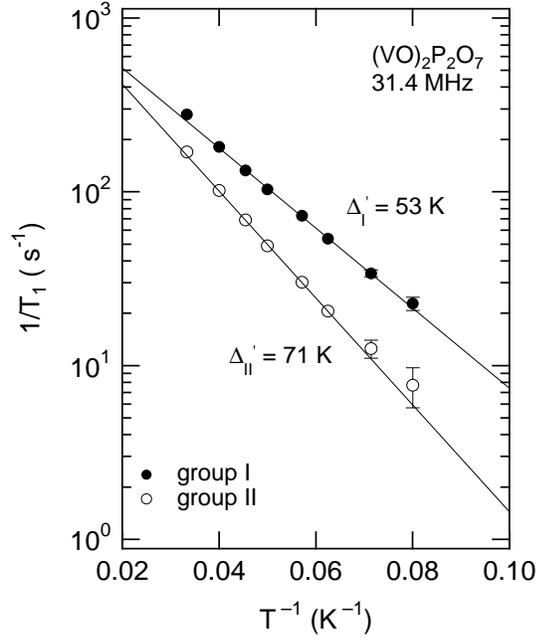}}
\medskip
\caption{
Temperature dependences of $1/T_1$ at the $^{31}$P sites 
belonging to the groups I and II.  Solid lines are the results of the 
least-squre fit of the data to the activation law $1/T_1 \propto \exp 
(-\Delta^\prime /T)$ with $\Delta^\prime$ of 53 K and 71 K for the 
groups I and II, respectively.}
\label{T1ofP}
\end{figure}

   The $T$ dependences of $1/T_1$ are shown in Fig.\ \ref{T1ofP}. 
It is clear that the $1/T_1$ for the different P groups show an activated 
behavior with different activation energies. By fitting the data to an 
activation form $1/T_1 \propto \exp(-\Delta^\prime /T)$, we obtained the 
gaps $\Delta^\prime_{\rm I}$ = 53 K and $\Delta^\prime_{\rm II}$ = 71 
K for the groups I and II, respectively.  The gap is larger for the 
group II which is consistent with the shift measurement.  It is also 
noted that the ratio of the two gap energies $\Delta^\prime_{\rm 
II}/\Delta^\prime_{\rm I}$ = 1.34 $\pm$ 0.10 for $1/T_1$ agrees 
roughly with that of the shifts $\Delta_{\rm II}/\Delta_{\rm I}$ = 
1.48 $\pm$ 0.10.  However, the absolute values of the gap determined 
from $1/T_1$ are larger than those obtained from the shifts as has 
been observed in other gapped quantum-spin systems.  
\cite{azuma94,shimizu95,takigawa96}

\subsection{$^{51}$V NMR}

   The results of the $^{31}$P NMR experiments show that 
(VO)$_2$P$_2$O$_7$ has two independent spin components with different energy 
gaps for the spin excitations.  This accounts for quite naturally the 
existence of two distinct excitation modes which are likely to come 
from the two inequivalent V chains running along the $b$ axis.  
However, one of the gap energies estimated from the shift does not 
agree with the one observed in the INS\cite{garrett97} and 
magnetization measurements \cite{yamauchi99} that needs further 
examinations.  This is partly because P nuclei feel transferred 
hyperfine fields from both spin components, which may result in the 
ambiguity in the estimation of the gap energy.  The $^{51}$V NMR is 
free from such an ambiguity because of a dominant on-site hyperfine 
coupling and is expected to give a more precise measure of the $T$ 
dependence of $\chi_{\rm spin}$ of each spin component.

   An intense free-induction-decay (FID) signal from V nuclei has been 
observed at low temperatures.  The frequency spectrum obtained from the 
Fourier transform of the FID signal is isotropic with FWHM of about 70 
kHz at 15.8 MHz and 4.2 K.  The FWHM is nearly $T$ independent and 
there is no indication of the line splitting up to 35 K.  We could not 
trace the signal above 35 K due both to the poor signal-to-noise ratio 
and to the shortening of the nuclear-spin relaxation time which 
prevents us from observing any transient signals by a pulsed NMR 
technique.

\begin{figure}
\centerline{\psfig{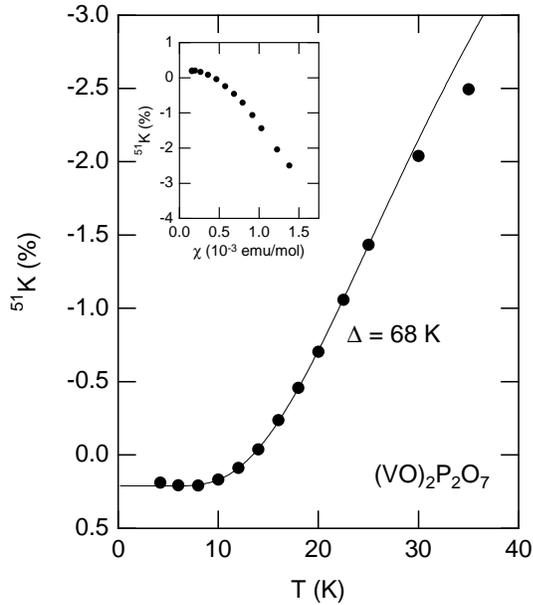}}
\medskip
\caption{
Temperature dependence of the shift $^{51}$$K$ at the $^{51}$V sites. Solid 
line is a fit of the data with a gap of 68 K (see text). Inset 
shows $^{51}$$K$ plotted against the bulk magnetic susceptibility.}
\label{K51vsT}
\end{figure}

   Figure\ \ref{K51vsT} shows the $T$ dependence of the shift 
$^{51}K$ at the peak position of the spectrum.  The shift is nearly 
constant below 10 K and varies strongly with temperature above 10 K.  In 
general, the shift $K$ of a nucleus belonging to the magnetic atom can 
be written as the sum of spin and orbital contributions, $K = K_{\rm 
spin}+K_{\rm orb}$.  By assuming that $\chi_{\rm spin}$ vanishes at 
$T$ = 0 owing to the spin gap, the orbital shift $K_{\rm orb}$ at V sites 
is estimated to be 0.212 $\pm$ 0.005 \%.  This is comparable with 
$K_{\rm orb}$ = 0.236 $\pm$ 0.005 \% in the reference material 
VOHPO$_4$$\cdot$${1\over 2}$H$_2$O with a similar local environment 
around V atoms.\cite{kikuchi98} The $T$ dependence of $K_{\rm spin}$ 
($=K-K_{\rm orb}$) is consistent with that of $\chi$ given in Eq.\ 
(\ref{Xspin}), and the energy gap for the V sites is determined to be 
68 $\pm$ 2 K by fitting $K_{\rm spin}$ to be proportional to $\chi$ in 
eq.\ (\ref{Xspin}).  This value is in excellent agreement with the gap 
of the second mode observed in the INS experiments.  The observation 
of the gap of the second mode via NMR suggests that the second mode is 
not a branch for the two-magnon bound states but is simply a 
one-magnon propagating mode.

   The inset of Fig.\ \ref{K51vsT} is a plot of $^{51}K$ versus 
bulk $\chi$ with temperature the implicit parameter.  It is obvious 
from the figure that the $^{51}K$ does not scale to the bulk $\chi$.  The 
$^{51}K$-$\chi$ plot concaves upward, which indicates that $\chi_{\rm 
spin}$ responsible for the observed $^{51}K$ decreases more rapidly 
with decreasing temperature than the bulk $\chi$.  The result of the 
$^{51}K$-$\chi$ plot also suggests that the bulk $\chi$ includes an 
additional contribution from the spin component with a gap smaller 
than that observed via $^{51}$V NMR.  Unfortunately, we could not 
detect a transient V signal corresponding to the spin component with a 
smaller gap energy, which results probably from the short nuclear 
spin-spin relaxation time $T_2$ of such V sites.

\begin{figure}
\centerline{\psfig{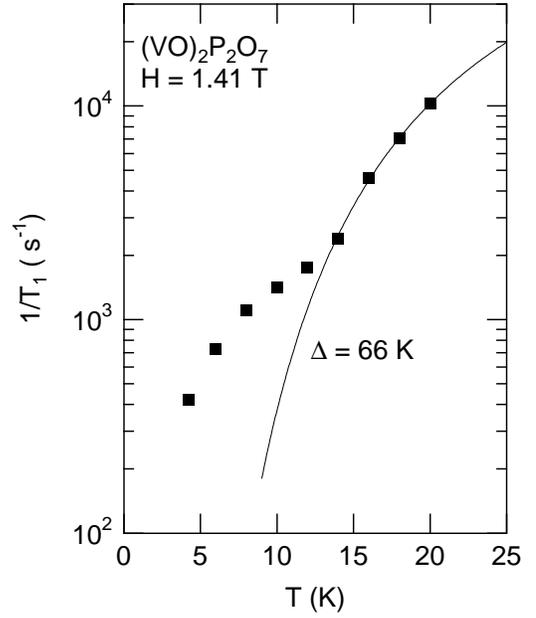}}
\medskip
\caption{
Temperature dependence of $1/T_1$ at the $^{51}$V sites. Solid line 
represents the activated temperature dependence with a gap of 66 K.}
\label{T1ofV}
\end{figure}

   The $T$ dependence of $1/T_1$ measured for the observed FID signal is shown 
in Fig.\ \ref{T1ofV}.  An activated behavior with a gap of 66 $\pm$ 2 K has 
been observed down to 14 K, below which the $1/T_1$ starts to deviate from the 
activation law probably because of extra relaxation processes.  The 
obtained gap for $1/T_1$ agrees quite well with that determined from 
the $^{51}K$ measurement, which is expected in the low-$T$ 
limit.\cite{troyer94} This is contrasted with the result at P sites 
where the gap energies for the shifts and $1/T_1$'s are different.  
The reason for this difference between V and P sites is not clear, 
although the different wave-vector dependence of the hyperfine form 
factor is likely to be the case.

\section{Discussion}
\label{sec:discussion}

   It is now evident from our NMR experiments that (VO)$_2$P$_2$O$_7$ is 
composed of two independent magnetic subsystems with different spin-gap 
energies. Since the different magnetic properties of the system generally come 
from the different crystallographic environment, it is reasonable to 
associate them with the two crystallographically-inequivalent V chains 
running along the $b$ axis which (VO)$_2$P$_2$O$_7$ inherently 
possesses.  The two-chain model gives consistent explanations for most 
of the present experimental results such as the observation of two 
distinct spin gaps (and $\chi_{\rm spin}$'s), the intensity ratio of 
the resonance lines for the two different P groups, and the 
underestimation of one of the spin-gap energies.  This model also 
accounts for the existence of the two distinct magnetic modes, which 
have been thought to be a one-magnon and the two-magnon bound states 
of the same alternating antiferromagnetic chain.

\subsection{Qualitative aspects}

   Before discussing the experimental results in detail, let us review the 
crystal structure of (VO)$_2$P$_2$O$_7$. As has been noted previously, 
(VO)$_2$P$_2$O$_7$ has eight inequivalent V and P sites, and 36 oxygen sites 
in the unit cell. This makes a crystallographic difference between the two 
kinds of V chains running along the $b$ axis.  The chain ``A'' consists of V 
atoms on V1-V4 sites and the chain ``B'' includes V5-V8 sites as shown in Fig.\ 
\ref{structure}.\cite{fnlabel} The V-V distances $d_{\rm VV}$'s in both 
the chains A and B alternate to give rise to the alternations of the 
exchange interactions, but the alternations in $d_{\rm VV}$'s are slightly 
different between the two chains.

   The difference of the spin gap and the resultant difference of 
$\chi_{\rm spin}$ between the chains can be understood qualitatively by 
considering that the different alternations in $d_{\rm VV}$ results in 
the different alternations in the exchange interaction as well.  This 
is very likely because the exchange interaction is expected to depend 
sensitively on the interatomic distance through the overlap of the 
related electronic orbitals.  In the alternating antiferromagnetic 
chains, the shape of the $\chi$-$T$ curve varies strongly with the 
ratio $\alpha = J_2/J_1$.\cite{bonner79} Therefore, if the chains A 
and B have different values of $\alpha$ owing to the different alternations 
in $d_{\rm VV}$, they exhibit different $T$ dependence of $\chi_{\rm 
spin}$'s.  The difference of the spin gap can also be explained 
because the spin gap of the alternating chain depends on the ratio 
$\alpha$ as well as the magnitude of $J_1$ and $J_2$.

   We did not mention explicitly the exchange pathways in the above argument. 
Garrett {\it et al}.  have pointed out that the exchange via the PO$_4$ 
complexes in the $b$ direction is essential in determining the magnetic 
properties of (VO)$_2$P$_2$O$_7$ as alternating antiferromagnetic 
chains.\cite{garrett97} Most of the present experimental results are 
consistent with such exchange pathways and are compatible with the 
two-chain model as described below.

\begin{figure}
\centerline{\psfig{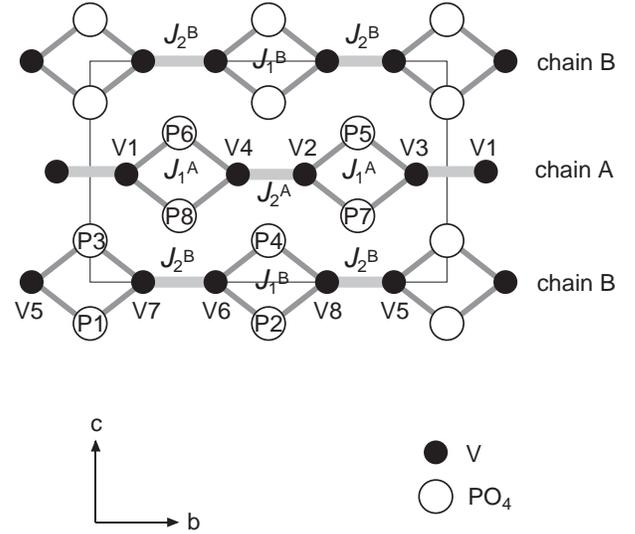}}
\medskip
\caption{
Schematic description of the exchange interactions in (VO)$_2$P$_2$O$_7$. 
The assignment of the inequivalent P sites to each alternating chain is 
also shown.  }
\label{exchange}
\end{figure}

   For the magnetic connection along the $b$ axis, two PO$_4$ 
complexes in between the VO$_5$ pyramids should be the important 
exchange pathways.  In the chain A, the exchange between V1 and V4 (V2 
and V3) sites is mediated by the PO$_4$ complexes containing P6 and 
P8 (P5 and P7) sites.  It includes the ÒdistantÓ V-O-P-O-V paths.  
There is also the exchange coupling between V4 and V2 (V3 and V1) sites via 
the shared VO$_5$ pyramidal edge having shorter V-O-V paths.  This is 
illustrated in Fig.\ \ref{exchange} together with the exchange paths 
for the chain B.  As a natural consequence of the exchange via PO$_4$ 
along the $b$ axis, eight inequivalent P sites are divided into two 
groups depending on to which chain they belong as the exchange paths: 
P5-P8 sites belonging to the chain A and P1-P4 sites belonging to the 
chain B.  This is consistent with the observation of two different P 
groups via $^{31}$P NMR.  In addition, since the number of the sites 
is the same for both P groups in this model, the intensity of 
the NMR lines should be the same.  This is what was observed in the 
$^{31}$P NMR experiments.

   The exchange coupling via the PO$_4$ complexes explains why the 
susceptibility of each spin component can be traced selectively by 
means of the $^{31}$P NMR.  In the two-chain model, the shift $K_{\rm 
A}$ at P sites in the chain A can be written as $K_{\rm A} = a_{\rm 
AA}\chi^{\rm A}_{\rm spin}+a_{\rm AB} \chi^{\rm B}_{\rm spin}$.  Here 
$\chi^{\rm A}_{\rm spin}$ and $\chi^{\rm B}_{\rm spin}$ the spin 
susceptibilities of the chains A and B; $a_{\rm AA}$ and $a_{\rm AB}$ 
the coupling constants at P sites in the chain A for the hyperfine 
fields transferred from the V sites in the chains A and B, 
respectively.  Similarly, for P sites in the chain B, the shift 
$K_{\rm B}$ can be written as $K_{\rm B} = a_{\rm BA}\chi^{\rm A}_{\rm 
spin}+a_{\rm BB}\chi^{\rm B}_{\rm spin}$ with $a_{\rm BA}$ and $a_{\rm 
BB}$ defined in a manner similar to $a_{\rm AA}$ and $a_{\rm AB}$.  
Since the transferred hyperfine field at nuclear sites in between the 
exchange-coupled magnetic atoms arises from the electron-spin transfer 
similar to that for the exchange, the hyperfine field at the P site is 
expected to come mainly from the V atoms in the same chain.  This 
means that $|a_{\rm AB}|$ and $|a_{\rm BA}|$ are much smaller than 
$|a_{\rm AA}|$ and $|a_{\rm BB}|$, indicating $K_{\rm A}$ and $K_{\rm 
B}$ to be roughly proportional to $\chi^{\rm A}_{\rm spin}$ and 
$\chi^{\rm B}_{\rm spin}$, respectively.

   It is expected that the interchain exchange coupling in the $c$ direction 
is weakened by the presence of the intrachain coupling via the PO$_4$ 
complexes.  This is because molecular orbitals of the PO$_4$ complex 
should be configured to connect the neighboring V atoms in the $b$ 
direction, which may be a microscopic origin of the dominant hyperfine 
coupling between the P nucleus and the V moments in the same chain.  As 
expected, the magnetic-excitation spectrum of (VO)$_2$P$_2$O$_7$ exhibits a 
very weak dispersion in the $c$ direction.\cite{garrett97} We notice here 
that the weak ferromagnetic coupling in the $a$ direction acts between 
the same kind of chain so that the chains A and B are magnetically decoupled.  
It is to be emphasized that the periodicity of the lattice is 
unchanged by the resultant grouping of V and P atoms: the magnetic 
unit cell is identical with the chemical one for both magnetic chains. 
Therefore, the spin-wave dispersion is observed in exactly the same 
Brillouin zone.  This is consistent with the INS results.

   Although the contribution from the different kind of chain to the 
hyperfine fields at P sites is considered to be small, it affects 
the estimation of the gap energies from $K_{\rm A}$ and $K_{\rm B}$.  This is 
serious for the P group belonging to the chain with a larger spin gap for the 
following reason.  Suppose that the spin gap $\Delta_{\rm B}$ of the 
chain B is smaller than the spin gap $\Delta_{\rm A}$ of the chain A.  
Then $\chi^{\rm B}_{\rm spin}$ starts to grow at lower temperatures 
than $\chi^{\rm A}_{\rm spin}$ and there may be a temperature region in 
which $\chi^{\rm B}_{\rm spin}$ grows rapidly with increasing temperatures 
while $\chi^{\rm A}_{\rm spin}$ is still vanishingly small.  In such a 
temperature region, $K_{\rm B}$ may be approximated by 
$a_{\rm BB}\chi^{\rm B}_{\rm spin}$ because both 
$a_{\rm AB}$ and $\chi^{\rm A}_{\rm spin}$ are small.  Hence the value 
of $\Delta_{\rm B}$ will be estimated correctly from $K_{\rm B}$.  On 
the other hand, for P sites in the chain A, the contribution from the 
different kind of chain $a_{\rm AB}\chi^{\rm B}_{\rm spin}$ may be comparable 
with, or greater than that from the same chain $a_{\rm AA}\chi^{\rm 
A}_{\rm spin}$.  As a result, $K_{\rm A}$ would start to increase at 
lower temperatures than $\chi^{\rm A}_{\rm spin}$.  This is inevitable 
as far as $a_{\rm AB}$ is finite and $\Delta_{\rm A}> \Delta_{\rm B}$, 
and will result in the underestimation of $\Delta_{\rm A}$.\cite{fnkb} 
In the present NMR experiment at P sites, the gap $\Delta_{\rm II}$ 
of the group II is smaller than the gap of the upper mode in the excitation 
spectrum, while $\Delta_{\rm I}$ agrees well with the gap of the lower 
mode.  In the light of the above argument, the underestimation of 
$\Delta_{\rm II}$ is considered to result from the small but finite 
contribution from V sites in the different kind of chain with a smaller gap.  
At this point, we may associate the P sites belonging to the group II (I) with 
the chain having a larger (smaller) spin gap.  The value of the gap determined 
from the $^{51}$V NMR experiment agrees well with the INS result and 
is considered to be more accurate for the larger spin-gap energy.

\subsection{Quantitative aspects}

   Most of the present experimental results can be explained qualitatively in 
terms of the two-chain model by taking account of the relatively strong V-V 
exchange via PO$_4$ in the $b$ direction.  The model is consistent also 
with the INS result and therefore seems to be appropriate for the description 
of the magnetic behavior of (VO)$_2$P$_2$O$_7$.  However, one crucial question 
arises here whether the small differences in $d_{\rm VV}$ can make the spin 
gaps of the two alternating chains be different by a factor of nearly two.  To 
answer this question, we estimated the exchange interactions in each 
alternating chain. It is found that the difference of the exchange is 
not very large and may be accounted for by the strong dependence of 
the exchange interaction on the interatomic 
distance.\cite{crawford76,kaneko87,massey90,millet98}

   For the analysis, we take the values of the spin gap to be 35 K and 68 K. 
The former is the one determined from the shift at P sites belonging to the 
group I, and the latter from the shift at V sites. Both are in excellent 
agreement with the zone-center gaps for the lower and upper modes in the 
spin-excitation spectrum, respectively.  The zone-boundary energy 
$\epsilon$$_{\rm ZB}$ taken from the spin-wave dispersion along the 
$b$ axis is 15.4 meV ($\approx$ 179 K) and is assumed to be the same for both 
chains.\cite{garrett97}  Then $J = (J_1+J_2)/2$ is calculated 
from the relation $\pi J/2 = \epsilon_{\rm ZB}$ to give the ratio 
$\Delta/J$, and the alternation parameter $\delta = 
(1-\alpha)/(1+\alpha)$ is determined by using the $\delta$ dependence 
of $\Delta/J$ presented by Uhrig and Schluz.\cite{uhrig96} From the 
values of $J$ and $\delta$, the exchange interactions $J_1$ and $J_2$ 
are calculated.  The results are summarized in Table \ref{table1}.

   We have assigned the chain A to have a larger exchange alternation (i.e., 
a larger value of $\delta$) resulting in a larger spin gap.  This is based on 
the following two assumptions: 1) the exchange via the PO$_4$ 
complexes is stronger than that through the shared VO$_5$ pyramidal 
edge, 2) the interaction is weaker for longer $d_{\rm VV}$ for both 
the exchange paths.  The V-V distance via the PO$_4$ complexes in 
the chain A is $\sim$5.14 {\AA}, which is shorter than that in the 
chain B of $\sim$5.16 {\AA}.\cite{fndistance} The largest exchange 
coupling ($J_1$ = 136 K) is therefore assigned to the V-O-P-O-V paths 
in the chain A.  The V-O-V path in the chain A is then assigned 
automatically to have the smallest coupling constant ($J_2$ = 92 K) 
without any additional assumptions.  The V-V distance is $\sim$0.02 
{\AA} longer than the corresponding $d_{\rm VV}$ in the chain B of 
$\sim$3.21 {\AA} and hence the assignment is self-consistent.  
Even if we regard the shared pyramidal edge as a dominant exchange 
pathway, the assignment can be made in the same way without 
inconsistency and now the spin gap of the chain A should be smaller 
than that of the chain B.

   It appeared that the exchange interactions of the chain A should differ 
from those of the chain B by $\sim$10 K ($\sim$10 \% in relative strengths) 
to give rise to the observed difference of the spin gap between the 
two chains.  Since the two-chain model is based on the crystallographic 
inequivalence of the chains, the difference of the exchange should be 
ascribed to the small differences ($\sim$0.02 {\AA} for both the 
exchange paths) in $d_{\rm VV}$.  With regard to the V-V exchange via 
the shared pyramidal edge, the difference may be attainable because in 
some copper salts with edge-sharing Cu-O bondings, the 0.02 {\AA} 
difference in the Cu-Cu distance gives rise to the difference in the 
exchange by $\sim$100 K.\cite{crawford76} The strong dependence of the 
exchange on the interatomic distance is also known empirically, 
\cite{kaneko87,massey90,millet98} putting the basis on the dependence 
of overlap integrals on distance.  For V-V pairs sharing common 
pyramidal edge, the empirical law predicts the exchange to decrease 
with increasing $d_{\rm VV}$ as ($d_{\rm VV})^{-10}$.\cite{millet98} 
The application to (VO)$_2$P$_2$O$_7$ yields $\sim$6 \% decrease of the V-V 
exchange on increasing $d_{\rm VV}$ from 3.21 to 3.23 {\AA}, which is 
comparable with the above estimation.  Although such an empirical 
relation is not known for V-V pairs connected by the PO$_4$ complexes 
because of the complicated and unresolved interaction pathways, it is 
plausible that the exchange is also sensitive to $d_{\rm VV}$ as the exchange 
via the common VO$_5$ edge.  The small difference in $d_{\rm VV}$'s between 
the two crystallographically-inequivalent chains is therefore 
considered to be crucial in making them magnetically distinct from 
each other.  Quantum chemical calculation of the exchange interaction via the 
PO$_4$ complexes is desirable for more thorough understanding of the 
magnetic properties of (VO)$_2$P$_2$O$_7$.

\section{Conclusion}
\label{sec:conclusion}

   We have presented the results of the extensive NMR study on the 
quasi one-dimensional gapped quantum antiferromagnet 
(VO)$_2$P$_2$O$_7$.  As revealed by the detailed line-shape analysis 
of the $^{31}$P NMR spectra and the measurement of $1/T_1$ at $^{31}$P 
sites, (VO)$_2$P$_2$O$_7$ has two independent spin components with different 
gap energies for the spin excitations.  The gap 
energies are determined from the temperature dependence of the shifts 
at $^{31}$P and $^{51}$V sites to be 35 K and 68 K, and are found to 
coincide with those observed in the INS experiments.

   The agreement of the gap energies between NMR and INS experiments suggests 
that the two magnetic modes in the excitation spectrum arise from the two 
distinct magnetic subsystems which (VO)$_2$P$_2$O$_7$ inherently possesses. 
We proposed that two crystallographically-inequivalent V chains running 
along the $b$ axis are magnetically distinct to contribute independently to 
the spin-excitation spectrum.  The two chains have different exchange 
alternations and spin gaps, which may result from the small difference in the 
V-V distances between the chains.  It is also suggested that the 
PO$_4$ complexes act as the important superexchange pathways.  This is 
essential in dividing the P sites into two groups and in observing the 
susceptibility of each alternating chain selectively via $^{31}$P 
NMR.  Quantitative estimates of the exchange interactions show that 
the difference of the exchange is not very large and may be attributed 
to the difference in the V-V distances between the two chains.

   To conclude, (VO)$_2$P$_2$O$_7$ is a complicated magnetic system containing 
two kinds of antiferromagnetic linear chains with different alternations in 
the exchange interaction. The energy for the spin excitations is different in 
each alternating chain, which is the origin of the two distinct magnetic modes 
in the excitation spectrum.

\acknowledgments
   This work was partly supported by a Grant-in-Aid for Scientific Research 
on Priority Areas from the Ministry of Education, Science, Sports and 
Culture.



\begin{table}
\caption{Spin gaps and exchange interactions in (VO)$_2$P$_2$O$_7$}
\label{table1}
\begin{tabular}{ccccc}
& $\Delta$ (K) & $J_1$ (K) & $J_2$ (K) & $J_2/J_1$\\
\tableline
chain A & 68 & 136 & 92 & 0.67\\
chain B & 35 & 124 & 103 & 0.83\\
\end{tabular}
\end{table}


\begin{references}
\bibitem{similarJ} 
This latter means that the exchange interaction is not very 
different for each spin component.
\bibitem{fnlabel} In numbering the sites, we followed the notation 
in ref.\ \onlinecite{nguyen95}.
\bibitem{fnkb} 
Of course, the contribution from V sites in the same 
chain becomes dominant at higher temperatures and then $K_{\rm A}$ 
traces $\chi^{\rm A}_{\rm spin}$.
\bibitem{fndistance} 
Strictly speaking, the V-V distances via PO$_{4}$ within the chain 
are slightly different: the V1-V4 and V2-V3 distances in the chain A are 
5.142 {\AA} and 5.137 {\AA}; the V5-V7 and V6-V8 distances 
in the chain B are are 5.157 {\AA} and 5.155 {\AA}, respectively. 
We disregarded these differences within the chain 
because they are relatively small compared with the differences between the 
chains and does not affect the assignment of the exchange to the V-V bond. 
The same is true for the V-V distances via the shared pyramidal edges.
\end{references}
\end{document}